\documentclass[11pt]{article}
\usepackage{graphics,graphicx}
\begin{document}

\title{Finite calculation of divergent
selfenergy diagrams}

\author{A.\, Aste, D.\, Trautmann\\
Institute for Theoretical Physics\\
Klingelbergstrasse 82, Basel, Switzerland}
\maketitle

\begin{abstract}
Using dispersive techniques, it is possible
to avoid ultraviolet divergences in the calculation of Feynman
diagrams, making subsequent regularization of
divergent diagrams unnecessary. We give a simple introduction to 
the most important features of such dispersive techniques
in the framework of the so-called finite
causal perturbation theory.
The method is also applied to
the 'divergent' general massive two-loop sunrise selfenergy diagram,
where it leads directly
to an analytic expression for the imaginary part of the
diagram in accordance with the literature,
whereas the real part can be obtained by a single integral
dispersion relation. It is pointed out that dispersive methods
have been known for decades and have been applied
to several nontrivial Feynman diagram calculations.\\
\vskip 0.1 cm
{\bf Keywords}: Causality, ultraviolet divergences\\
\vskip 0.1 cm
{\bf PACS}: 11.10.-z, 11.15.Bt, 12.20.Ds, 12.38.Bx.
\end{abstract}

\newpage

\section{Introduction}
In quantum field theory, one usually starts from
fields and a Lagrangean which describes the interaction.
These objects get quantized and S-matrix elements or Greens functions
are then constructed by the help of Feynman rules.
However, these Feynman rules lead to products of distributions
with singular behavior. This leads to the well-known
ultraviolet-divergences in perturbation
theory. The occurrence of such divergences has led to the
invention of several ingenious formal procedures like
Pauli-Villars regularization \cite{Pauli} or
dimensional regularization \cite{Dim}.
The situation remained not entirely satisfactory for several
decades. So Feynman in his Nobel lecture remarked:
"I think that the renormalization
theory is simply a way to sweep the difficulties
of the divergences of quantum electrodynamics under the rug."

Today, the situation is not so bad as Feynman described it.
The regularization procedures which remove infinities from the
theory are well understood and it has become clear that UV
divergences do not imply an inconsistency of the theory
in general, but are a consequence of our current
perturbative formulation of quantum field theories.
Renormalization-group methods also shed a new light
on their interpretation.

It is still widely unknown that it is possible to
treat Feynman diagrams in such a manner that 
ultraviolet (UV) divergences do not appear in the
calculations. This can be achieved by making explicit use of the
causal structure of quantum field theory, i.e. by using
dispersive techniques. Contrary to widespread belief, the use of
dispersion relations in practical calculations is not unusual
and has been employed by various authors in phenomenologically
important evaluations of
calculationally demanding quantum electrodynamic corrections
such as higher-order binding corrections to the
two-loop bound-state quantum electrodynamic self-energy.
Artru {\em{et al.}} have used dispersive techniques
as early as in 1966 for a leading-order calculation of the Lamb shift
\cite{Artru,Abarbanel}. Pachucki has employed these relations
for the evaluation of higher-order binding corrections to the two-loop
Lamb shift \cite{Pachucki}, and mentioned in his paper that the ultraviolet
divergences canceled automatically in the calculations.

A very natural and rigorous approach to axiomatic perturbation theory
which has been applied successfully to all relevant interactions of the
Standard model
\cite{appl1,appl2,appl0,appl3} and supersymmetric models
\cite{appl4,appl5}
was provided by a classic paper of
H. Epstein and V. Glaser in 1971 \cite{eg,eg2}.
Their method, called
finite causal perturbation theory or FCPT for short in this paper,
avoids UV divergences from the start by
defining mathematically correct time-ordered products for distributions.
The resulting distributions are
smeared out by test functions, avoiding provisionally volume and infrared
divergences at once.
An extensive introduction to the causal method can be found in the
textbook of G. Scharf on finite quantum electrodynamics \cite{GScharf}.
There is now also a growing interest in FCPT, because it allows to formulate
a consistent renormalization theory on a curved physical background
\cite{fredenhagen}.

Basically, FCPT provides a clearly defined strategy for the
dispersive calculation of Feynman diagrams. It renders all diagrams
finite, but does not automatically solve problems related
to the specification
of normalization conditions and possible anomalies
arising in the regularization procedure, which may violate
symmetries which are necessary to give physical meaning.
These problems like e.g. the formulation of Ward identities
or the introduction of interacting fields in the framework
of FCPT have been trated extensively in recent years
\cite{dufre1,dufre2,hurth0,hurth1,hurth2}.

In FCPT,
the ansatz for the $S$-matrix as a power 
series in the coupling constant is crucial, namely $S$ is considered as a sum of 
smeared operator-valued distributions
\begin{equation}
S(g)={\bf 1}+\sum_{n=1}^{\infty}\frac{1}{n!} \int\!\! d^{4}x_{1}
\ldots d^{4}x_{n}\, T_{n}(x_{1},\ldots, x_{n})\,g(x_{1})\cdot\ldots\cdot 
g(x_{n})\quad  ,
\end{equation}
where the  Schwartz test function $g\in\cal{S}(\bf{R}^{4})$ switches the 
interaction and provides the infrared cutoff.
The basic formulation of causality in FCPT, which had already been
used by Bogoliubov {\em{et al.}} \cite{bogol}, is
\begin{equation}
S(g_1+g_2)=S(g_2)S(g_1) \quad , \label{causality1}
\end{equation}
if the support of $g_2(x)$ is later than $\mbox{supp} \, g_1$ in some
Lorentz frame
(usually denoted by $\mbox{supp} \, g_2 > \mbox{supp} \, g_1$).
The condition allows the construction of
the $n$-point distributions $T_n$ as a  well-defined  `renormalized'
time-ordered product expressed 
in terms of Wick monomials of free fields $:\!{\cal O}(x_{1},\ldots,x_{n})$:
\begin{equation}
T_{n}(x_{1},\ldots, x_{n})=\sum_{ { \cal{ O}}} :\!{\cal 
O}(x_{1},\ldots,x_{n})\!:\,
t_{n}^{{ \cal O}}(x_{1}-x_{n},\ldots,x_{n-1}-x_{n})\,. \label{smatrix}
\end{equation}
where 
the $t_{n}^{{ \cal O}}$ are C-number distributions.
$T_{n}$ is constructed inductively from 
the first order $T_{1}(x)$,
which describes the interaction among the quantum fields,  and from 
the lower orders $T_{j}$, $j=2,\ldots,n-1$ by means of Poincar\'e
covariance and causality.
The inductive construction of the n-point distributions $T_n$
can be considered
as the main technical difference of the theory to
other approaches, since all lower orders
$T_1,...T_{n-1}$ must be calculated first in order to construct $T_n$.
But the commonly chosen form of the $T_n$
\begin{displaymath}
T_n(x_1,...,x_n)=
\end{displaymath}
\begin{equation}
\sum_{\pi} \Theta(x_{\pi_1}^0-x_{\pi_2}^0)\cdot ...
\cdot \Theta(x_{\pi_{n-1}}^0-x_{\pi_n}^0) T_1(x_{\pi_1})\cdot ... \cdot
T_1(x_{\pi_n})
\end{equation}
where the sum runs over all $n!$ permutations,
is not an unambiguous definition, since it contains the product of Heaviside
distributions with other singular distributions. This leads to the
well-known UV divergences in the calculation of Feynman diagrams.
One can illustrate this fact by the following simple 
one-dimensional example.
The product of a Heaviside
distribution $\Theta(t)$ with
a $\delta$-distribution $\delta(t)$ or even its
derivative $\delta'(t)$ is obviously not well-defined.
The Fourier transforms of the distributions are
\begin{equation}
\hat{\Theta}(\omega)=-\frac{i}{\sqrt{2 \pi}} \frac{1}{\omega-i0} \quad ,
\quad \hat{\delta'}(\omega)=\frac{i\omega}{\sqrt{2 \pi}} \quad .
\end{equation}
The ill-defined product $(\Theta \delta')(t)$ goes over into a 
non-existing convolution by Fourier transform
\begin{equation}
(\Theta\delta')(t) \rightarrow \frac{1}{2 \pi} (\hat\Theta \ast
\delta')(\omega)=\frac{1}{(2 \pi)^2} \int_{-\infty}^{+\infty}
d\omega' \, \frac{\omega-\omega'}
{\omega'-i0}
\end{equation}
which is definitely 'UV divergent'.

It is the aim of this paper to demonstrate how UV finite
results are obtained in the framework of FCPT by avoiding
problematic Feynman integrals.
This is done in Section 2, where technical details are explained.
In Section 3 the method is applied to the well-known
introductory case of a scalar selfenergy diagram, and
in section 4, we apply the method to the less trivial
case of the two-loop sunrise diagram with arbitrary masses.

\section{The method}

We start our considerations with a scalar neutral massive field $\phi_m$
with mass $m$ which satisfies the free Klein-Gordon
equation. It can be decomposed into a negative and positive frequency
part according to
\begin{displaymath}
\phi_m(x)=\phi_m^-(x)+\phi_m^+(x)=
\end{displaymath}
\begin{equation}
(2 \pi)^{-3/2} \int \frac{d^3k}{\sqrt{2 E}} \Bigl[ a(\vec{k})
e^{-ikx}+a^\dagger(\vec{k}) e^{ikx} \Bigr] ,
\end{equation}
where $E=\sqrt{\vec{k}^2+m^2}$.
The commutation relations for such a field are given by the Jordan-Pauli
distributions
\begin{equation}
[\phi_m^{\mp}(x) , \, \phi_m^{\pm}(y)]=-i D_m^\pm(x-y) \, ,
\label{commutator}
\end{equation}
which have the Fourier transforms
\begin{equation}
\hat{D}_m^{\pm}(p)=(2\pi)^{-2} \int d^4x \, D_m^{\pm}(x) e^{ipx}=
\pm \frac{i}{2 \pi} \Theta(\pm p^0) \delta(p^2-m^2) .
\end{equation}
Since the commutator
\begin{equation}
[\phi_m(x),\phi_m(y)]=-iD_m^-(x-y)-iD_m^+(x-y)=:
-iD_m(x-y)
\end{equation}
vanishes for spacelike distances $(x-y)^2 < 0$
due to the requirement of microcausality, it
follows immediately that the Jordan-Pauli distribution
$D_m$ has {\emph{causal support}}, i.e. it vanishes outside the closed
forward and backward lightcone
\begin{equation}
\mbox{supp} \, D_m(x) \subseteq \overline{V}^- \cup \overline{V}^+ ,
\end{equation}
\begin{equation}
\overline{V}^\pm=\{x \, | \, x^2 \ge 0, \, \pm x^0 \ge 0 \}.
\end{equation}
A crucial observation is the fact that the retarded propagator is
given in real space by the covariant formula
\begin{equation}
D_m^{ret}(x)=\Theta(x^0) D_m(x) ,
\end{equation}
which goes over into a convolution in momentum space
\begin{equation}
\hat{D}_m^{ret}(p)=(2 \pi)^{-2} \int d^4 k \, \hat{D}_m(k) \hat{\Theta}
(p-k) . \label{convo}
\end{equation}
The Heaviside distribution $\Theta(x^0)$ could be replaced by
$\Theta(vx)$ with an arbitrary
vector $v \in V^+$ inside the forward lightcone.
The Fourier transform of the Heaviside distribution $\Theta(x^0)$
can be easily calculated
\begin{equation}
\hat{\Theta}(p)= (2 \pi)^{-2} \lim_{\epsilon \rightarrow 0}
\int d^4x \, \Theta(x^0) e^{-\epsilon x^0} e^{ip_0x^0-i \vec{p}\vec{x}}=
\frac{2 \pi i}{p^0+i0} \delta^{(3)}(\vec{p}) .
\end{equation} 
For the special case where $p$ is in the forward lightcone $\overline{V}^+$,
we can go to a Lorentz frame where $p=(p^0,\vec{0})$
($=p^0=p_0$ as a shorthand) such that eq.
(\ref{convo}) becomes
\begin{equation}
\hat{D}_m^{ret}(p^0)=\frac{i}{2 \pi} \int dk^0 \frac{\hat{D}_m
(k^0)}{p^0-k^0+i0}=
\frac{i}{2 \pi} \int dt \, \frac{\hat{D}_m
(tp^0)}{1-t+i0} .
\end{equation}
For arbitrary $p \in V^+$, $\hat{D}_m^{ret}$ is therefore
given by the {\emph{dispersion relation}}
\begin{equation}
\hat{D}_m^{ret}(p)=\frac{i}{2 \pi} \int dt \, \frac{\hat{D}_m
(tp)}{1-t+i0} . \label{disprel}
\end{equation}
It is trivial but
instructive to perform the actual calculation of $\hat{D}_m^{ret}$.
Exploiting the $\delta$-distribution in
\begin{equation}
\hat{D}_m(p)=\frac{i}{2 \pi} \mbox{sgn}(p^0) \delta(p^2-m^2),
\end{equation}
we obtain
\begin{displaymath}
\hat{D}_m^{ret}(p)=-\frac{1}{(2 \pi)^2} \int dt \, \frac{
\mbox{sgn}(tp^0) \delta(t^2p^2-m^2)}{1-t+i0}=
\end{displaymath}
\begin{displaymath}
-\frac{1}{(2 \pi)^2} \int dt \, \frac{
\Bigl[ \delta(t-\frac{m}{\sqrt{p^2}})- \delta(t+\frac{m}{\sqrt{p^2}})
\Bigr]}{2 \sqrt{p^2} m (1-t+i0) }=
\end{displaymath}
\begin{equation}
-\frac{1}{(2 \pi)^2} \frac{1}{p^2-m^2+i0}.
\end{equation}
We recover the analytic expression for the Feynman propagator,
which coincides with
\begin{equation}
\hat{D}_m^{ret}(p)=\hat{D}_m^F(p)+D_m^-(p)=
-\frac{1}{(2 \pi)^2} \frac{1}{p^2-m^2+ip_0 0}
\end{equation}
for $p \in V^+$.
The imaginary part of the Feynman propagator is given by
\begin{equation}
\mbox{Im} \, (\hat{D}_m^F(p))=\frac{i}{4 \pi} \delta(p^2-m^2)
\end{equation}
and can be deduced directly from the causal
Jordan-Pauli distribution $\hat{D}_m$ in an obvious way.

\subsection{The scalar one-loop selfenergy diagram}
As a first step, we define the causal
distribution $d(x-y)$ in conformity with (\ref{commutator})
as the following vacuum expectation value of a massive and
a massless field
(the colons denote normal ordering) 
\begin{equation}
(-i) \, d(x-y):=
\langle 0 | [:\phi_{m}(x)\phi_o(x):,
:\phi_{m}(y)\phi_o(y):] | 0 \rangle
\end{equation}
which has again causal support due to microcausality.
It is sufficient to consider
\begin{displaymath}
(-i) \, r(x-y):=
-\langle 0 | :\phi_{m}(y)\phi_o(y): \,
:\phi_{m}(x)\phi_o(x): | 0 \rangle=
\end{displaymath}
\begin{equation}
D_{m}^-(x-y)D_o^-(x-y),
\end{equation}
since $d(x-y)=r(x-y)-r(y-x)$.
The product of the two Jordan-Pauli distributions in real space
goes over into a convolution in momentum space
\begin{equation}
\hat{r}(p)=\frac{i}{(2 \pi)^2} (\hat{D}^-_{m} \ast
\hat{D}^-_o)(p), \label{conv2}
\end{equation}
such that $\hat{d}(p)=\hat{r}(p)-\hat{r}(-p)$.

Since we do {\emph{not}} calculate the
{\emph{time-ordered}} product of fields,
the Feynman propagators are replaced by Jordan-Pauli distributions.

The integral appearing in (\ref{conv2}) is readily evaluated
if one exploits the $\delta$-distribution stemming from the massless
field contraction in
\begin{equation}
\hat{r}(p)=
-\frac{i}{(2 \pi)^4} \int d^4q \, \Theta(-q^0) \delta(q^2)
\Theta(q^0-p^0) \delta((p-q)^2-m^2). \label{self1}
\end{equation}
$\hat{r}(p)$ vanishes outside the closed backward lightcone
due to Lorentz invariance and
the two $\Theta$-distributions in eq. (\ref{self1}).
Therefore we can go to a Lorentz frame where $p=(p^0<0, \vec{0})$,
and using the abbreviation $E=\sqrt(\vec{q}^{\, 2}+m_1^2)$ leads to
\begin{displaymath}
\hat{r}(p^0)=
-\frac{i}{(2 \pi)^4} \int \frac{d^3 q}{2E}
\delta(p_0^2+2p^0E-m^2)\Theta(-E-p^0)=
\end{displaymath}
\begin{displaymath}
-\frac{i}{2(2 \pi)^3 |p^0|} \int \, d |\vec{q}| |\vec{q}|
\delta\Bigl(\frac{p^0}{2}+E-\frac{m^2}{2p^0}\Bigr)
\Theta(-E-p^0)=
\end{displaymath}
\begin{equation}
-\frac{i}{2(2 \pi)^3 |p^0|} \Theta(-p^0) \Theta(p_0^2-m^2)
\sqrt{\Bigl(\frac{p^0}{2}-\frac{m^2}{2 p^0}\Bigr)^2},
\end{equation}
and for arbitrary $p$ we get the intermediate result
\begin{equation}
\hat{r}(p)=-\frac{i}{32 \pi^3} \frac{p^2-m^2}{p^2}
\Theta(-p^0) \Theta(p^2-m^2).
\end{equation}
Naive use of
the dispersion relation (\ref{disprel}) for the real part
of the sunrise diagram would lead to an ultraviolet divergent
expression. But it can be shown \cite{GScharf} that the finite
part of the diagram is given for $p \in V^+$
by a subtracted dispersion relation
(which is also called {\em{splitting formula}})
\begin{equation}
\hat{t}^{\, ret}(p)=\frac{i}{2 \pi} \int
\limits_{-\infty}^{\infty} dt \, \frac{\hat{d}
(tp)}{(t-i0)^{\omega+1}(1-t+i0)} , \label{suptrel}
\end{equation}
where $\omega$ is the power counting degree of the diagram,
which is in our case $\omega=0$ since
the diagram is logarithmically divergent.
$\hat{t}^{\, ret}(p)$ is normalized such that
all derivatives of order $\leq \omega$ of $\hat{t}^{\, ret}$ vanish at
$p=0$, i.e. in the present case
\begin{equation}
\hat{t}^{\, ret}(0)=0.
\end{equation}
These statements are only meaningful if the
derivatives of order $\leq \omega$ of $\hat{t}^{\, ret}(p)$
exist in the sense of ordinary functions. This is
assured in most cases
by the existence of a massive field in the theory.

We mention here that eq. (\ref{suptrel}) corresponds to the
dispersion relation (113,10) used in the familiar textbook by
Landau and Lifshitz on quantum electrodynamics \cite{Landau}.

Roughly speaking, the additional term in the denominator of
(\ref{suptrel}) has a simple explanation. Writing (\ref{suptrel})
for $p=(p^0,\vec{0})$ as
\begin{equation}
\hat{t}^{\, ret}(p^0)=\frac{i}{2 \pi} (p^0)^{\omega+1}
\int \limits_{-\infty}^{\infty} dk^0 \, 
\frac{\hat{d}(k^0)}{(k^0-i0)^{\omega+1}(p^0-k^0+i0)},
\end{equation}
it becomes obvious that the division by $(k^0)^{\omega+1}$
acts as a kind of inverse differentiation on the causal
distribution $d(x)$ in real space.
The transformed distribution has a less
critical scaling behavior, and can be multiplied in a
well-defined way by the time-ordering $\Theta$-distribution.
After the splitting, the original part of $d(x)$
in the forward lightcone is recovered
by differentiating $\omega+1$ times along the
time axis, corresponding to a multiplication
of the distribution by $(p^0)^{\omega+1}$ in momentum space.
This observation also shows the connection of the method to
differential renormalization \cite{prange}.
For further mathematical details we refer also to
\cite{lazzarini}.

The splitting integral can be evaluated by elementary methods
\begin{equation}
\hat{t}^{\, ret}(p)=
-\frac{1}{4(2 \pi)^4} \int \limits_{-\infty}^{\infty} dt \,
\frac{\mbox{sgn}(t) \Theta(t^2p^2-m^2)(t^2p^2-m^2)}{t^3(1-t+i0)p^2},
\quad p \in V^+
\end{equation}
and leads directly to 
\begin{equation}
\hat{t}(p)=
\frac{1}{4(2 \pi)^4} \Biggl[ \frac{m^2-p^2}{p^2} \log \Bigl(
\frac{m^2-p^2-i0}{m^2} \Bigr) +1 \Biggr].
\end{equation}
Distribution theory leaves the freedom to add a constant to
the result, which corresponds to a local term $c_o \delta(x)$
in real space. Such local terms have to be restricted
by further symmetry considerations in practical cases.

We arrived thus on a very direct way to a finite expression
for the scalar one-loop selfenergy diagram, which corresponds
up to a prefactor to
the regularized expression of the Feynman integral
\begin{equation}
\int \frac{d^4 k}{(k^2+i0)((p-k)^2-m^2+i0)}.
\end{equation}

We finally mention that one-loop diagrams in two space-time
dimensions are treated  in the framework of FCPT in \cite{Walther},
whereas three-dimensional QED is treated in \cite{GScharf}.

\section{The Sunrise Diagram}
In the following section,
we will utilize the strategy discussed above
for the calculation of the imaginary
part of the sunrise diagram. This will show that the causal
treatment of Feynman diagrams is also applicable to non-trivial cases.
An extensive introduction to the calculation of two-loop diagrams
in the causal method
has been given in \cite{Aste}, but the case of the sunrise diagram
is missing there.
Due to the fundamental structure of the sunrise diagram,
it is clear that it has been investigated by the use of many
different approaches, and in this paper it serves only as a
convenient example for a dispersive calculation.
We give therefore a short overview over the literature
which is related to the subject in the following.

There is now an increased interest in precise calculations
of multi-loop Feynman diagrams. In the general massive case,
relevant for future high precision calculations in the electroweak theory
which go beyond the currently available one-loop electroweak 
radiative corrections \cite{DennerPozzorini}, 
the number of parameters makes it impossible to obtain
results in the usual 
analytic form already in the case of the two-loop sunrise selfenergy 
diagram shown in Fig.1. 
When one or two internal particles are massless, the four-dimensional
results can be obtained in terms of dilogarithms
\cite{Tausk,Kallen,Broad,Fleischer}.
The situation is more complicated when all the
virtual particles in the diagram have (different) masses. Such a
situation occurs e.g. in the two-loop off-shell contributions
to the Higgs selfenergy. In three dimensions, the expression for
the sunrise diagram is quite compact even for different masses
\cite{Rajantie},
but in the four dimensions, there are arguments \cite{ScharfDiss}
that the result can not be expressed in terms of polylogarithms or other
well-known special functions.

There is nowadays also a widely accepted procedure of expressing
radiative correction calculations in terms of a limited number of 
master integrals (MI) \cite{TkaChet}, which reduces 
the problem of finding analytic expressions for two-loop diagrams
to the careful determination of the MI. 
The method has also the advantage that, with a correct bookkeeping of 
the recurrence relations arising from integration by parts identities, 
the MI of a given problem can be reused in 
more complicated calculations.

As already mentioned, the analytical calculation of MI (in terms of the usual 
polylogarithms and their generalizations) is possible 
only in cases with high symmetry, i.e.
when the number of different scales (internal masses 
and external momenta 
or Mandelstam variables) is small like in QCD calculations, where 
all masses are set to zero, or in QED-type cases, where only the 
electron mass is different from zero, or when the external variables
are fixed to particular values (zero or mass shell condition). 
Another possibility of help in analytic calculations is sometimes 
offered by the exploitation of particular simplifying conditions, like the 
smallness of some ratios of the parameters allowing the corresponding 
expansion.  
\begin{figure}[htb]
        \centering
        \includegraphics[width=7.5 cm]{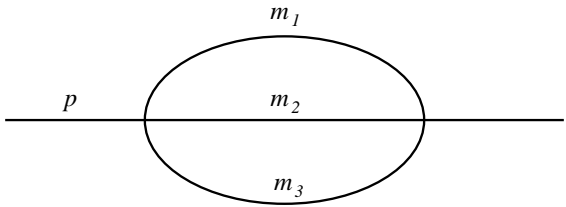}
        \caption{The general massive two-loop sunrise selfenergy diagram.}
        \label{sunrise.eps}
\end{figure}

The MI of the sunrise diagram were 
recognized to be expressible in closed form as a combination of four 
Lauricella functions, a special class of generalized hypergeometric 
series \cite{BeSch} (and references therein). 
The method provides efficient multiple series expansions for the regions
of small $|p^2|$, i.e. $|p^2| < \mbox{max}(m_i^2)$, and of large 
$|p^2| \gg (m_1+m_2+m_3)^2$, but some problems arise in the
intermediate region.

There were also efforts devoted to the investigation of properties 
in special points (i.e. at $p^2 = 0, \infty$, pseudothresholds 
and threshold).
The analytical expansions of the MI at $0$ and $\infty$ are given in 
\cite{BeSch} and \cite{CaffoLaporta}; the values at pseudothresholds and 
threshold in \cite{Berends};
the analytical expansions at pseudothresholds can be found in \cite{Caffo1};  
a semi-analytical expansion at threshold in \cite{DavySmi} and also in 
configuration space technique in \cite{Groote}, while the 
complete analytical expansions at threshold are presented  
in \cite{Caffo2}. 

For numerical evaluation purposes, it is possible to cast 
the general massive selfenergy diagram in 
a double integral representation  
and in the particular case of the sunrise diagram in a single integral 
representation \cite{BeSch,PostTausk,Aste,bauberger}
(and references therein).
The configuration space technique is also exploited in the numerical 
approach \cite{BeSch}, \cite{GrootePi}.
In a recent approach rearrangements of the integrand, driven by the 
Bernstein-Tkachov theorem, are introduced to improve numerical 
convergence \cite{Passarino}. 
A different and interesting method is the use of the recurrence relations
as difference equations to numerically evaluate the MI \cite{Laporta}.

It is of general interest to provide the widest possible
range of insight into the structure and analytic properties of
multi-loop diagrams.
In the present paper we exploit the causal structure of
the sunrise diagram, which immediately leads to a straightforward
evaluation of the imaginary part of the diagram by elliptic
functions. The real part of the diagram can then be obtained from a single
integral dispersion relation. The dispersive method works especially
well for diagrams which depend only on one external momentum, although
its has already been used for different kinds of cases.

As in the case of the one-loop selfenergy diagram, we define the causal
distribution $d(x-y)$ according to eq. (\ref{commutator})
as the vacuum expectation value of three massive fields
(the colons denote normal ordering) 
\begin{displaymath}
(-i) \, d(x-y):=
\end{displaymath}
\begin{equation}
\langle 0 | [:\phi_{m_1}(x)\phi_{m_2}(x)\phi_{m_3}(x):,
:\phi_{m_1}(y)\phi_{m_2}(y)\phi_{m_3}(y):] | 0 \rangle
\end{equation}
which has causal support.
As for the one-loop case, we consider
\begin{displaymath}
(-i) \, r(x-y):=
\end{displaymath}
\begin{displaymath}
-\langle 0 | :\phi_{m_1}(y)\phi_{m_2}(y)\phi_{m_3}(y): \,
:\phi_{m_1}(x)\phi_{m_2}(x)\phi_{m_3}(x): | 0 \rangle
\end{displaymath}
\begin{equation}
=i D_{m_1}^-(x-y)D_{m_2}^-(x-y)D_{m_3}^-(x-y),
\end{equation}
since $d(x-y)=r(x-y)-r(y-x)$.
The product of the three Jordan-Pauli distributions in real space
goes over into a threefold convolution in momentum space
\begin{equation}
\hat{r}(p)=-\frac{1}{(2 \pi)^4} (\hat{D}^-_{m_1} \ast
\hat{D}^-_{m_2} \ast \hat{D}^-_{m_3})(p), \label{conv3}
\end{equation}
such that $\hat{d}(p)=\hat{r}(p)-\hat{r}(-p)$.

The dispersion relation (\ref{disprel}) for the real part
of the sunrise diagram
is given by the splitting formula
($p \in V^+$)
\begin{equation}
\hat{t}^{\, ret}(p)=\frac{i}{2 \pi} \int dt \, \frac{\hat{d}
(tp)}{(t-i0)^{3}(1-t+i0)} , \label{suptrel2}
\end{equation}
since the power counting degree of the quadratically
divergent diagram is in this case $\omega=2$.
$\hat{t}^{\, ret}(p)$ is normalized such that
all derivatives of order $\leq \omega$ of $\hat{t}^{\, ret}$ vanish at
$p=0$
\begin{equation}
\hat{t}^{\, ret}(0)=0, \quad
\frac{\partial}{\partial p_\mu} \hat{t}^{\, ret}(0)=0, \quad
\frac{\partial^2}
{\partial p_\mu \partial p_\nu} \hat{t}^{\, ret}(0)=0.
\end{equation}

\subsection{The Imaginary Part}
We will now calculate the three-fold convolution of the Jordan-Pauli
distributions in momentum space given by eq. (\ref{conv3}).
First we evaluate
\begin{displaymath}
\hat{d}_{12}(p)=(\hat{D}^-_{m_1} \ast \hat{D}^-_{m_2})(p)=
\end{displaymath}
\begin{equation}
-\frac{1}{(2 \pi)^2} \int d^4q \, \Theta(-q^0) \delta(q^2-m_1^2)
\Theta(q^0-p^0) \delta((p-q)^2-m_2^2). \label{step1}
\end{equation}
$\hat{d}_{12}(p)$ vanishes outside the closed backward lightcone
due to Lorentz invariance and
the two $\Theta$-distributions in eq. (\ref{step1}).
In a Lorentz frame where $p=(p^0<0, \vec{0})$ we obtain
($E=\sqrt(\vec{q}^{\, 2}+m_1^2)$)
\begin{displaymath}
\hat{d}_{12}(p^0)=
\end{displaymath}
\begin{displaymath}
-\frac{1}{(2 \pi)^2} \int \frac{d^3 q}{2E}
\delta(p_0^2+2p^0E+m_1^2-m_2^2)\Theta(-E-p^0)=
\end{displaymath}
\begin{displaymath}
-\frac{1}{2(2 \pi) |p^0|} \int \, d |\vec{q}| |\vec{q}|
\delta\Bigl(\frac{p^0}{2}+E+\frac{m_1^2-m_2^2}{2p^0}\Bigr)
\Theta(-E-p^0)=
\end{displaymath}
\begin{displaymath}
-\frac{1}{2(2 \pi) |p^0|} \Theta(-p^0) \Theta(p_0^2-(m_1+m_2)^2)
\end{displaymath}
\begin{equation}
\sqrt{\Bigl(\frac{p^0}{2}+\frac{m_1^2-m_2^2}{2 p^0}\Bigr)^2-m_1^2},
\end{equation}
and for arbitrary $p$ we get the intermediate result
\begin{displaymath}
\hat{d}_{12}(p)=-\frac{1}{8 \pi p^2} \Theta(-p^0) \Theta(p^2-(m_1+m_2)^2)
\end{displaymath}
\begin{equation}
\sqrt{p^4-2(m_1^2+m_2^2)p^2+(m_1^2-m_2^2)^2},
\end{equation}
which is symmetric in $m_1,m_2$ and exhibits the typical two-particle
threshold behavior. Applying (\ref{suptrel}) with $\omega=0$ to
$\hat{d}_{12}(p)$ would generate the retarded one-loop propagator for
a diagram with masses $m_1$ and $m_2$.
As a next step we calculate
\begin{displaymath}
\hat{r}(p)=-\frac{1}{(2 \pi)^4} (\hat{D}_{m_3}^- \ast \hat{d}_{12})(p)=
-\frac{i}{4(2 \pi)^6} \int d^4q 
\end{displaymath}
\begin{displaymath}
\Theta(-q^0) \delta(q^2-m_3^2)
\Theta(q^0-p^0) \Theta((p-q)^2-(m_1+m_2)^2)
\end{displaymath}
\begin{equation}
\frac{\sqrt{(p-q)^4-2(m_1^2+m_2^2)(p-q)^2+(m_1^2-m_2^2)^2}}
{(p-q)^2} \quad .
\end{equation}
For $p=(p^0,\vec{0})$ we obtain
$\Bigl(E=\sqrt{\vec{q}^{\, 2}+m_3^2} \Bigr)$
\begin{displaymath}
\hat{r}(p)=-\frac{i}{4(2 \pi)^6} \int \frac{d^3q}{2 E}
\end{displaymath}
\begin{displaymath}
\frac{\Theta(-E-p^0) \Theta(p_0^2+2p_0E+m_3^2-(m_1+m_2)^2)}
{p_0^2+2p_0E+m_3^2} \, I(p_0)=
\end{displaymath}
\begin{displaymath}
-\frac{i}{4(2 \pi)^5} \Theta(-p^0) \Theta(p_0^2-(m_1+m_2+m_3)^2)
\end{displaymath}
\begin{equation}
\int \limits_{m_3}^{-\frac{p^0}{2}+\frac{(m_1+m_2)^2-m_3^2}{2p^0}}
dE \, \frac{\sqrt{E^2-m_3^2}}{p_0^2+2p_0E+m_3^2} \, I(p_0) ,
\label{monster}
\end{equation}
\begin{displaymath}
I(p_0)=\Bigl(
(p_0^2+2p_0E+m_3^2)^2-
\end{displaymath}
\begin{equation}
2(m_1^2+m_2^2)(p_0^2+2p_0E+m_3^2)+
(m_1^2-m_2^2)^2\Bigr)^{1/2}.
\end{equation}
The integral (\ref{monster}) can be written in a compact manner
if we substitute $s=p_0^2+2p_0E+m_3^2$. Going back to arbitrary
Lorentz frames by replacing $p^0$ by $-\sqrt{p^2}$ where necessary, we
obtain the following representation of $\hat{r}(p)$:
\begin{equation}
\hat{r}(p)=-\frac{i}{16(2 \pi)^5 p^2} \Theta(-p^0)
\Theta(p^2-(m_1+m_2+m_3)^2) \, \mbox{J}(p),
\end{equation}
where
\begin{equation}
\mbox{J}(p)=\int \limits_{s_1}^{s_4}
\frac{ds}{s} \sqrt{(s-s_1)(s-s_2)(s_3-s)(s_4-s)},
\label{ellint}
\end{equation}
and the variables $s_2<s_1<s_4<s_3$ are defined via
\begin{displaymath}
s_1=(m_1+m_2)^2, \quad s_2=(m_1-m_2)^2,
\end{displaymath}
\begin{equation}
s_3=(\sqrt{p^2}+m_3)^2, \quad s_4=(\sqrt{p^2}-m_3)^2.
\end{equation}
The integral in (\ref{ellint}) can be expressed by complete elliptic
integrals $E$, $K$ and $\Pi$:
\begin{equation}
\mbox{J}=
\gamma_1E(\alpha)+\gamma_2K(\alpha)
+\gamma_3\Pi (\beta_1,\alpha)
+\gamma_4 \Pi(\beta_2,\alpha), \label{result}
\end{equation}
where
\begin{displaymath}
\alpha=\sqrt{\frac{(s_4-s_1)(s_3-s_2)}{(s_3-s_1)(s_4-s_2)}},
\end{displaymath}
\begin{displaymath}
\beta_1=\frac{s_4-s_1}{s_4-s_2},
\quad \beta_2=\frac{s_2(s_4-s_1)}{s_1(s_4-s_2)},
\end{displaymath}
\begin{displaymath}
\gamma_1=\frac{1}{4}(\sum \limits_{i} s_i)\sqrt{(s_3-s_1)(s_4-s_2)},
\end{displaymath}
\begin{displaymath}
\gamma_2=\frac{1}{4} \sqrt{\frac{s_4-s_2}{s_3-s_1}}
(s_1-s_2)(s_1-s_2+3s_3+s_4),
\end{displaymath}
\begin{displaymath}
\gamma_3=-\frac{1}{4}\frac{(s_1-s_2)}{\sqrt{(s_4-s_2)(s_3-s_1)}}
(\sum \limits_{i} s_i^2-2 \sum \limits_{i<j} s_is_j),
\end{displaymath}
\begin{equation}
\gamma_4=-2\frac{s_4 s_3 (s_1-s_2)}{\sqrt{(s_4-s_2)(s_3-s_1)}}.
\end{equation}
For the sake of convenience and clarity, we note the
definition of the complete elliptic integrals:
\begin{displaymath}
E(k)=\int \limits_{0}^{1} dt \, \sqrt{\frac{1-k^2t^2}{1-t^2}} \, ,
\end{displaymath}
\begin{displaymath}
K(k)=\int \limits_{0}^{1} \frac{dt}{\sqrt{(1-t^2)(1-k^2t^2)}} \, ,
\end{displaymath}
\begin{equation}
\Pi(\nu,k)=\int \limits_{0}^{1} \frac{dt}{\sqrt{(1-t^2)(1-k^2t^2)
(1-\nu^2t^2)}} \, .
\end{equation}
$E$, $K$ and $\Pi$ satisfy very many identities, therefore it might be
possible that an even more compact form can be found for J.
The analytic expression for $\hat{d}(p)$ follows from $\hat{r}(p)$ simply
by replacing $\Theta(-p^0)$ by $-\mbox{sgn}(p^0)$, and the imaginary
part of the sunrise diagram is given by
\begin{equation}
\mbox{Im} \, (\hat{t}(p))=\frac{i}{32 (2 \pi)^5 p^2}
\Theta(p^2-(m_1+m_2+m_3)^2) \, \mbox{J} \quad .
\end{equation}
The result is indeed fully symmetric under permutations of $m_1,m_2,m_3$,
although this is not obvious from eq. (\ref{result}).
But for the higher symmetry cases, J becomes quite simple. For
$m_1=0$, $m_2=m_3=m$, J reduces to
\begin{equation}
\mbox{J}(p)=\frac{1}{2} (p^2+2m^2)\sqrt{p^2(p^2-4m^2)}, \label{mass2}
\end{equation}
for $m_1=m_2=0$, $m_3=m$ we have
\begin{equation}
\mbox{J}(p)=\frac{1}{2} (p^2-m^2)(p^2+m^2), \label{mass1}
\end{equation}
and for $m_1=m_2=m_3=0$ , J collapses to
\begin{equation}
\mbox{J}(p)=\frac{1}{2} p^4. \label{massless}
\end{equation}
Applying eq. (\ref{suptrel2}) to (\ref{mass2}) and (\ref{mass1})
leads to the results mentioned in the introduction which are expressible
by polylogarithms and associated functions. But the subtracted dispersion
relation (\ref{suptrel2}) does not work in the totally massless case,
where one encounters an additional infrared divergence which requires
additional regularization that can be derived also as a limit
from the massive case.
The solution in the massless case, relevant e.g. for QCD or
quantum gravity, is
\begin{displaymath}
\hat{t}^{\, ret}(p^2)=\hat{t}(p^2)+\hat{r}(p)=
\end{displaymath}
\begin{equation}
-\frac{1}{32 (2 \pi)^6} p^2 \log
\frac{-(p^2+ip_0 0)}{M^2},
\end{equation}
\begin{displaymath}
\hat{t}(p^2)=-\frac{1}{32 (2 \pi)^6} p^2 \log \frac{-(p^2+i0)}{M^2}=
\end{displaymath}
\begin{equation}
-\frac{1}{32 (2 \pi)^6} p^2 \Bigl(\log \Bigl| \frac{p^2}{M^2} \Bigr|
-i \pi \Theta(p^2)\Bigr) \label{masslesst}
\end{equation}
with a real scaling parameter M. Eq. (\ref{masslesst}) reflects
also the high energy behavior of the massive propagator $\hat{t}
\sim p^2 \mbox{log}(p^2)$.
The prefactor $(2 \pi)^{-6}$ in the expressions above is due to the
fact that we have included a prefactor $(2 \pi)^{-2}$ in our
definition of the Feynman propagator.

\section{Final remarks and conclusion}
A closed analytic expression has been derived for the
imaginary part of the scalar sunrise diagram with arbitrary masses
of the virtual particles.
The technical difference between the usual Feynman diagram
calculations and the method used in the present paper is that
the occurring integrals are reduced step by step to single
integrals by utilizing the $\delta$-distributions in the
Jordan-Pauli distributions instead of performing a Wick rotation.
The real part of the sunrise diagram can
be expressed for all values of the external impulse by a single
integral, because the causal distribution $\hat{d}$ which is related
to the imaginary part of the diagram has the special form
\begin{equation}
\hat{d}(p)=h(p^2) \mbox{sgn}(p^0) \Theta(p^2-m_{tot}^2),
\quad
m_{tot}=m_1+m_2+m_3,
\end{equation}
therefore the dispersion relation (\ref{suptrel2}) can
also be written for $p \in V^+$
\begin{displaymath}
\hat{r}^{ret}(p)=\hat{t}(p)+\hat{r}(p)=
\end{displaymath}
\begin{equation}
\frac{i}{2 \pi} (p^2)^{[\omega/2]+1}
\int \limits_{m_{tot}^2}^{\infty} ds \, \frac{h(s)}
{s^{[\omega/2]+1}(p^2-s+ip_0 0)},
\end{equation}
where $[\omega/2]$ denotes the largest integer $\leq \omega/2$.
This formula can be extended to space-like $p$ by analytic
continuation
\begin{equation}
\hat{r}^{\, ret}(p^2)=\frac{i}{2 \pi} (-p^2)^{[\omega/2]+1}
\int \limits_{m_{tot}^2}^{\infty} ds \, \frac{h(s)}
{s^{[\omega/2]+1}(-p^2-s)},
\end{equation}
such that that an access to numerical and analytic investigations
is provided for the whole range of external momenta.

The problem of overlapping
divergences is automatically solved by the causal method due to the
inductive construction of the theory, whereas in usual
renormalization schemes it becomes a nontrivial problem to show
that all sub-diagrams of a given diagram can be renormalized in
a consistent way. In the context of the BPHZ method, the problem
was solved by the forest formula, an algorithm which disentangles
all divergences in subdiagrams \cite{Zimmermann}.
This is one of the conceptual strengths of the causal approach.
Furthermore it should be pointed out that on the level of the sunrise
diagram discussed in this paper, the calculation of the imaginary
part can as well be understood by the Cutkoski rules. But at higher orders,
one should clearly distinguish between these rules which
rely mainly on the unitarity of the theory, whereas in the causal
approach the main input is causality.

We have used the example of the surise diagram in order to illustrate
in a pedagogical manner the strategy and fundamental properties
of the dispersive approach for the calculation of Feynman diagrams.
We refer also to the recent literature on the specific case of the
sunrise diagram, where also non-trivial vertex structures are treated
by dispersive and non-dispersive methods \cite{bauberger,delb1,delb2}.

\end{document}